\documentclass[10pt,twocolumn,letterpaper]{article}

\usepackage[pagenumbers]{cvpr}

\usepackage[T1]{fontenc} 

\usepackage{cuted}      
\usepackage{currfile}
\usepackage{caption}
\usepackage{booktabs}
\usepackage{multirow}
\usepackage{pifont}
\usepackage{xspace}
\usepackage{amsthm}
\usepackage{siunitx}
\usepackage[table]{xcolor}
\definecolor{cafeblue}{RGB}{234,244,250}
\usepackage{algorithm}
\usepackage{algpseudocode}

\newcommand{\method}{CaFe\xspace}        
\newcommand{\dataname}{WOMD-PACT\xspace} 

\newcommand{\cmark}{\ding{51}}
\newcommand{\xmark}{{\color{red!75!black}\ding{55}}}

\newtheorem{definition}{Definition}
\newtheorem{proposition}{Proposition}

\definecolor{cvprblue}{rgb}{0.21,0.49,0.74}
\usepackage[breaklinks,colorlinks,allcolors=cvprblue]{hyperref}

\def\confName{CVPR}
\def\confYear{2027}

\title{CaFe: Causally Faithful Conditional Traffic Scenario Generation}

\author{
Steven Yu Tung\textsuperscript{1,2,*} \quad
Shubham Vaijanath Phoolari\textsuperscript{1,3,*} \quad
Christoph Lauer\textsuperscript{1} \\
Abhinav Valada\textsuperscript{2} \quad
Steven Peters\textsuperscript{3} \\[0.7em]
{\small
\textsuperscript{1}CARIAD SE, Germany \quad
\textsuperscript{2}University of Freiburg, Germany \quad
\textsuperscript{3}TU Darmstadt, Germany
} \\[0.3em]
{\small \textsuperscript{*}Equal contribution.}
}

\begin{document}

\maketitle

\begin{strip}
\centering
\includegraphics[width=\linewidth]{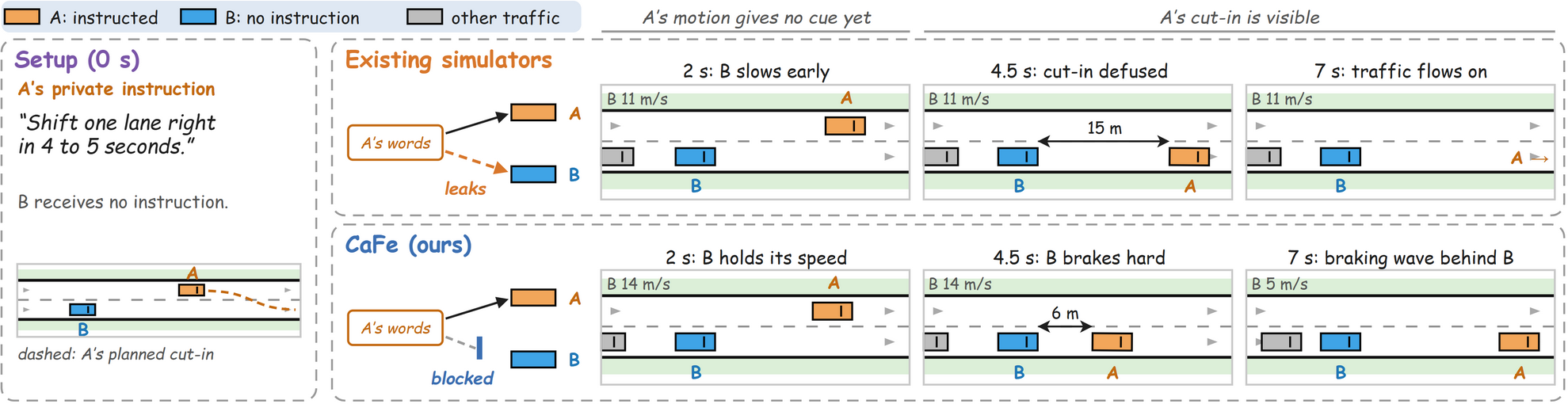}
\captionsetup{hypcap=false}
\captionof{figure}{
\textbf{Intent leakage and how CaFe prevents it.}
Agent $A$ is privately instructed to cut in late ahead of $B$,
which receives no instruction.
\emph{Top (existing simulators):}
$A$'s intent reaches $B$ through shared conditioning or
unexecuted future trajectories. $B$ slows before the cut-in
is observable, opening a large gap and defusing the intended
interaction.
\emph{Bottom (CaFe):}
We separate private intent from public beliefs derived from
observed motion. $B$ maintains speed until $A$ cuts in,
then brakes, triggering a braking wave in surrounding traffic.
CaFe preserves the intended interaction without leaking
private intent. Trajectories and speeds are schematic.
}
\captionsetup{hypcap=true}
\label{fig:teaser}%
\end{strip}

\begin{abstract}
Language-conditioned traffic simulation enables scalable test scenarios steered by natural language. The goal is to control selected agents while surrounding traffic reacts to their observed behaviour. For example, a vehicle instructed to cut in late should cause the following vehicle to brake only when the manoeuvre becomes observable. Yet simulators can leak private instructions across agents, causing surrounding vehicles to react before the instructed behaviour becomes observable. Such premature reactions defuse desired events and alter the ensuing closed-loop scenario. Existing benchmarks measure what agents do, not what they could have known, leaving this failure unmeasured. We formalise it as \emph{intent leakage} and address it through evaluation, data and modelling. Our paired intervention test holds an instructed agent's executed trajectory fixed while varying its private instruction, isolating leakage-induced changes in surrounding agents. We also construct \dataname{}, an automatically generated, verifiable dataset of 531,099 WOMD scenes with timed behaviour tags with reaction episodes identified through a counterfactual \emph{ghost test} and grounded private instructions. Finally, we introduce \method{}, a dual-stream diffusion-forcing simulator fine-tuned in closed loop that restricts inter-agent interactions to public beliefs, preserving legitimate anticipation without leaking private intent. Our evaluation reveals that existing conditioning designs displace 23--83\% of nearby uninstructed agents through private-intent leakage, even when the instructed agent's executed motion is held fixed. \method{} eliminates measurable leakage while maintaining comparable WOSAC realism (0.763), demonstrating that causal faithfulness can be enforced without sacrificing measured simulation realism.
\end{abstract}


\section{Introduction}
\label{sec}

The validation of autonomous vehicles (AVs) requires, among many other aspects, exposing the system to safety-critical interactions, such as aggressive cut-ins or failures to yield, which occur rarely in logged driving data. Traffic simulation addresses this scarcity by reproducing such interactions on demand. Rule-based simulators and scenario description languages~\cite{asam2020openscenario,dosovitskiy2017carla,fremont2019scenic} offer precise control but require scenarios to be authored by hand, whereas data-driven simulators~\cite{suo2021trafficsim,xu2023bits,wu2024smart,jiang2024scenediffuser} learn realistic behaviour from logged driving yet provide little control over which interactions arise. Conditional generation reconciles the two by steering a learned behaviour model with goal positions~\cite{zhong2023ctg,zhou2025nexus}, behaviour tags~\cite{tan2024prosim} or natural language~\cite{tan2023lctgen,zhong2023ctgpp,tan2024prosim,xia2024interacttraj,chang2025langtraj}. Instructions are typically given to a subset of agents, while the others respond based on what they observe. For instance, one vehicle may be instructed to cut in ahead of another, leaving the response of surrounding traffic to the simulator.

This expectation does not hold in current simulators. ProSim~\cite{tan2024prosim} conditions each agent's policy on the text of the entire scene, while LangTraj~\cite{chang2025langtraj} additionally lets agents attend to the text-conditioned future trajectories of others. We term this exposure \emph{intent leakage}. Its visible consequence is a \emph{preemptive reaction}, where an uninstructed agent responds to an intention before any motion expresses it. In the example above, the surrounding vehicle decelerates before the instructed vehicle begins its cut-in (\cref{fig:teaser}). The conflict is thereby resolved in advance, and the premature response propagates through the closed-loop rollout.

Standard evaluation is insensitive to intent leakage. Realism
benchmarks compare generated rollouts with recorded
ones~\cite{montali2023wosac}, irrespective of the information
on which each agent acts. A simulator that reveals a manoeuvre
in advance may reproduce the recorded response while allowing
surrounding agents to anticipate it prematurely. Interactive
motion prediction can legitimately condition on another agent's
future trajectory~\cite{tolstaya2021cbp,seff2023motionlm}.
In closed-loop simulation, however, that future is not yet
observable to the other agents.

We formalise intent leakage as any influence of an instruction
on another agent that does not pass through
the instructed agent's observable motion. This extends interventional
evaluation of interactive prediction to private instructions
held by any agent. Measuring leakage requires separating it from legitimate reaction, since removing an instruction also changes the instructed agent's motion, to which others may rightly respond. We therefore compare paired closed-loop rollouts with shared random seeds, replaying the instructed agents' generated trajectories with and without their instructions. The uninstructed agents observe identical instructed-agent motion in both runs, so any difference in their responses reveals leakage, while a simulator without leakage yields exactly zero.

Preventing leakage requires care in both the data and the model. An instruction that describes another agent's intended behaviour reveals its plan before that behaviour becomes observable. Existing language datasets often describe a reaction together with its cause~\cite{tan2024prosim,chang2025langtraj}. \dataname{} therefore derives private instructions from measured behaviour. Timed behaviour tags record what each road user does and when, while a counterfactual \emph{ghost test} identifies the likely causes of reactions. At evaluation, reacting agents receive no instruction and must respond to observed motion.

CaFe then separates an agent's intent from what others can perceive,
while aiming to preserve the underlying simulator's capabilities.
A private stream conditions each agent on its
own instruction, while a language-blind public stream predicts
all agents from shared history and executed motion.
In inter-agent attention, agents read only public beliefs
about one another, excluding intent leakage by construction
while preserving anticipation from observed motion.

\noindent Our key contributions are:
\begin{itemize}
  \item \textbf{Intent leakage and counterfactual evaluation.}
  We identify and formalise intent leakage in
  language-conditioned traffic simulators and introduce
  a paired closed-loop replay test that isolates it from
  legitimate reactions by keeping executed motion fixed
  while varying the private instruction.

  \item \textbf{\dataname: Private, verifiable traffic instructions.}
  We introduce \dataname, a large-scale dataset built
  from 531,099 WOMD scenes, with private, fact-grounded
  and temporally aligned instructions across interacting
  road users, together with directed reaction annotations.

  \item \textbf{\method: Causally faithful traffic generation.}
  We introduce \method, a dual-stream language-conditioned
  simulator that separates private intent from public
  predictive beliefs. It replaces access to other agents'
  private future representations with beliefs derived
  from observed motion, preventing direct intent leakage
  while preserving anticipation and reaction to
  surrounding traffic.
\end{itemize}

\section{Related Work}
\label{sec:related}

\noindent\textbf{Scenario generation.}
Scenario languages~\cite{asam2020openscenario,dosovitskiy2017carla,fremont2019scenic}
script behaviour and triggers by hand.
Learned traffic models~\cite{suo2021trafficsim,xu2023bits,wu2024smart,huang2026vbd,jiang2024scenediffuser,tan2025scenediffuserpp,niedoba2023djinn}
imitate logged behaviour, judged by distributional
realism~\cite{montali2023wosac}.
Controllable generators steer them with
guidance~\cite{zhong2023ctg},
scenario tokens~\cite{pronovost2023scenariodiffusion},
latent scene layouts~\cite{chitta2024sledge,rowe2025scenariodreamer}
or goal states.
Nexus~\cite{zhou2025nexus} decouples token noise so
goals and observed history can be held clean while
the rest is generated.
Our backbone follows MDG's~\cite{huang2025mdg} scene encoding backbone
and VBD's~\cite{huang2026vbd}
query-centric relation encoding.

\par\noindent\textbf{Language-conditioned generation.}
LCTGen~\cite{tan2023lctgen} and
InteractTraj~\cite{xia2024interacttraj}
map text through an LLM to structured codes that
decode the whole scene.
CTG++~\cite{zhong2023ctgpp} and
LD-Scene~\cite{peng2025ldscene}
let an LLM write guidance for scene-level diffusion.
ChatScene~\cite{zhang2024chatscene} and
Text2Scenario~\cite{cai2025text2scenario}
emit simulator programs.
ProSim~\cite{tan2024prosim} and
LangTraj~\cite{chang2025langtraj}
train directly on text, using either a scene-level
prompt read by every agent's policy or a shared
sentence with role tags.
ECoSim~\cite{chen2026ecosim} keeps a pretrained simulator frozen and trains lightweight FiLM control adapters on its decoder layers for sketch, latent and text control.
These works ask whether the instructed agent obeys.
None asks who else can read the instruction.

\par\noindent\textbf{Language datasets for driving.}
ProSim-Instruct-520k~\cite{tan2024prosim}
turns rule-based tags into scene-level prompts
with an LLM.
InterDrive~\cite{chang2025langtraj}
adds human interaction labels paraphrased into
shared sentences.
WOMD-Reasoning~\cite{li2025womdreasoning}
generates question--answer pairs from rule-based facts.
Instruction sets centred on the ego vehicle or a
focal agent~\cite{felemban2026imotion,roy2026doplan,roy2024doscenes,deruyttere2019talk2car,wu2025nuprompt}
and captioning and QA
corpora~\cite{kim2018bddx,malla2023drama,sima2024drivelm,arai2025covla}
address other tasks.
None provides instructions that are private to one
agent, available for every interacting agent and
traceable to measured facts (Tab.~\ref{tab:datasets}).
\begin{table}[t]
\centering
\small
\setlength{\tabcolsep}{3.2pt}
\newcommand{\pmark}{$\sim$}
\resizebox{\linewidth}{!}{%
\begin{tabular}{@{}llccccc@{}}
\toprule
Dataset & Base & Verif. & Private & Symm. & Inter. & Timed \\
\midrule
ProSim-Instruct~\cite{tan2024prosim} & WOMD & \xmark & \xmark & \cmark & \xmark & \pmark \\
InterDrive~\cite{chang2025langtraj} & WOMD, nuPlan & \pmark & \xmark & \pmark & \cmark & \xmark \\
WOMD-Reas.~\cite{li2025womdreasoning} & WOMD & \pmark & \xmark & \xmark & \pmark & \pmark \\
InstructWaymo~\cite{felemban2026imotion} & WOMD & \cmark & -- & \xmark & \xmark & \xmark \\
doPlan~\cite{roy2026doplan} & nuPlan & \cmark & -- & \xmark & \xmark & \pmark \\
\midrule
\dataname{} (ours) & WOMD & \cmark & \cmark & \cmark & \cmark & \cmark \\
\bottomrule
\end{tabular}}
\caption{\textbf{Instruction datasets for trajectory-level driving.}
We compare factual verification, private per-agent instructions, symmetric coverage, interaction identification and timing at behaviour onset. \pmark{} indicates partial support and -- denotes single-agent or inapplicable settings.}
\label{tab:datasets}
\end{table}

\par\noindent\textbf{Interaction labels and leakage.}
WOMD marks interacting pairs without
direction~\cite{ettinger2021womd}.
CausalAgents~\cite{roelofs2024causalagents}
and InterDrive~\cite{chang2025langtraj}
collect human labels.
M2I~\cite{sun2022m2i} and
TrafficGraphNet~\cite{kumar2021trafficgraph}
infer who yields from arrival order.
Our ghost test instead asks what a car would have
done had nobody been there.
In joint forecasting, conditioning one agent on
another's future yields ``reactions'' that precede
their cause~\cite{seff2023motionlm}.
Conditional predictors do so by
design~\cite{rowe2023fjmp,tolstaya2021cbp}.
This is appropriate for a forecasting query but
not for a simulated agent.
Learning from what text happens to mention is
a known shortcut~\cite{goyal2017vqav2,xie2025drivebench,dehaan2019causal}.
We bring these concerns to language-conditioned
generation and make them measurable.

\section{Problem Formulation}
\label{sec:leakage}
\subsection{Conditional scenario generation}

A conditional scenario generator rolls a logged traffic scene forward
in closed loop. A test engineer steers some agents with instructions;
all agents are simulated, and the others react only to what they
observe. At time $t$, the context
\begin{equation}
c^t = \big(m,\ \ell^{\le t},\ \{x_j^{\le t}\}_{j=1}^{N}\big)
\label{eq:context}
\end{equation}
holds the map $m$, the traffic-signal states $\ell^{\le t}$ and the
executed histories of all $N$ agents, where $x_j^t$ is the state of
agent $j$ at time $t$. Each instructed agent $i$ receives a private
instruction $C_i$, a set of clauses; uninstructed agents have
$C_i=\emptyset$. With $\mathcal{C}=\{C_i\}_{i=1}^{N}$, the generator
samples the next motion of all agents every $\Delta$ seconds from
\begin{equation}
p\big(x_{1:N}^{t+\Delta}\mid c^t,\ \mathcal{C}\big)
\label{eq:generator}
\end{equation}
and executes it, which updates the context to $c^{t+\Delta}$.
Everything derived from $C_i$, including agent $i$'s planned but
unexecuted motion, must stay private: other agents may react to what
$i$ has executed, but not to its instruction or intent.




\begin{figure}[t]
\centering
\includegraphics[width=\linewidth]{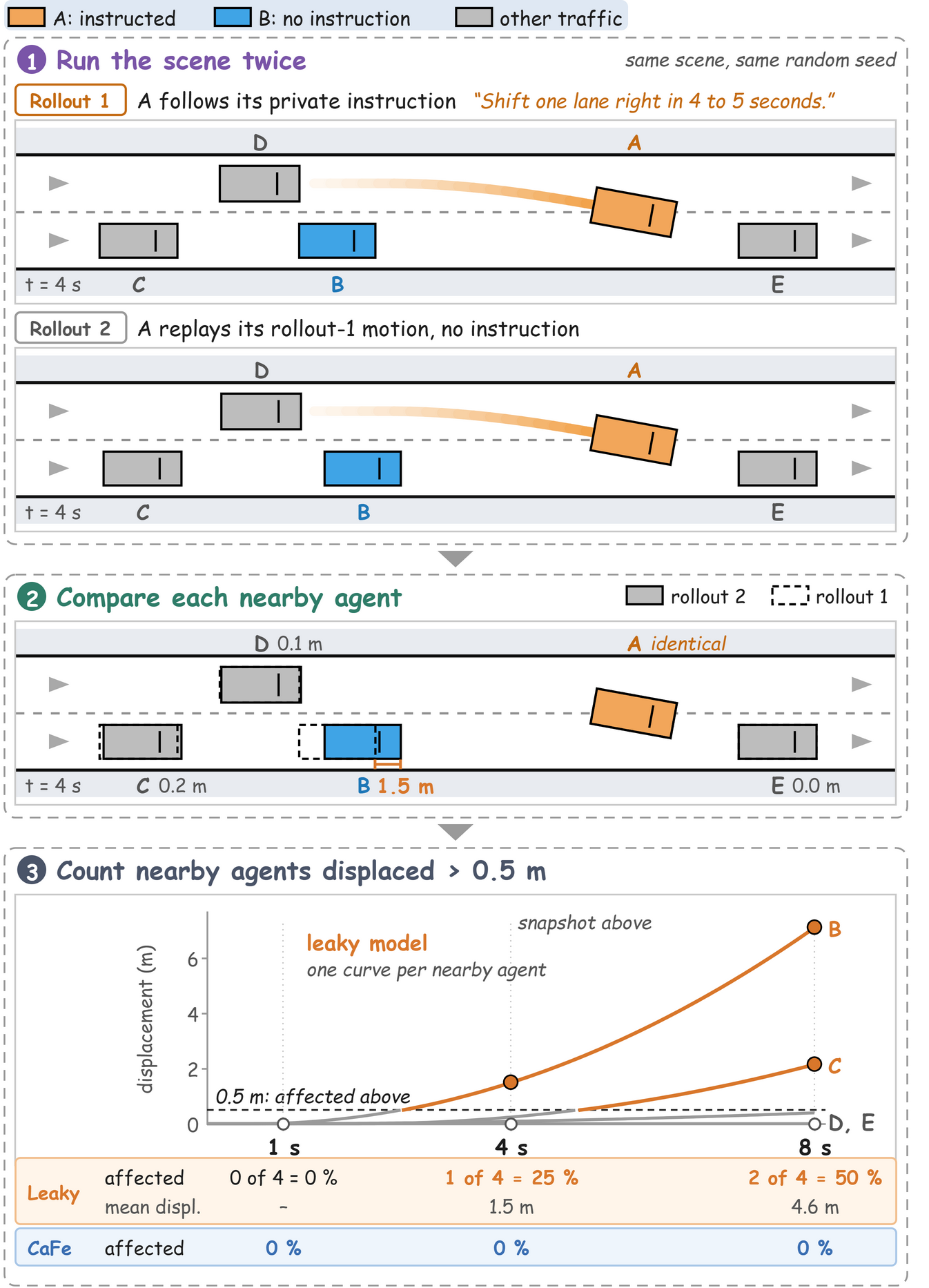}
\caption{
\textbf{Measuring intent leakage.}
We replay the instructed agent’s motion without its private instruction, compare surrounding agents across rollouts, and quantify leakage along two dimensions: the fraction of affected agents and their mean displacement.
}
\label{fig:metric}
\end{figure}

\subsection{Intent leakage and causal faithfulness}
\label{sec:faithful}

An agent should not react to information that only another agent
possesses. We formalise this as causal faithfulness.

\begin{definition}[Causal faithfulness]
\label{def:faithful}
Let $\xi$ denote the randomness the generator draws, independent of
$\mathcal{C}$. A generator is causally faithful
if, for every agent $i$ and step $t$,
\begin{equation}
p\big(x_i^{t+\Delta}\mid c^t,\xi,\mathcal{C}\big)
= p\big(x_i^{t+\Delta}\mid c^t,\xi,C_i\big).
\label{eq:faithful}
\end{equation}
Any violation of \eqref{eq:faithful} is \emph{intent leakage}.
\end{definition}

Other agents' instructions can thus reach agent $i$ only through
executed motion in $c^t$. The definition concerns instructions only:
randomness shared across agents, through which joint models
coordinate, is permitted as long as it carries no instruction. Nor
does it forbid anticipation. If $A$ slows down or lines up for a turn, that
is already in $c^t$, and $B$ may and should use it. What $B$ cannot
use is what exists only in $A$'s instruction or its unexecuted plan.

\subsection{Measuring intent leakage}
\label{sec:metric}

Leakage is hard to identify in a single rollout because an early
brake can look realistic. We isolate it with a paired intervention
(Fig.~\ref{fig:metric}). We first roll out a scene with the
instructions $\mathcal{C}$. We then roll it out again with all
instructions removed and the same randomness $\xi$, while every
instructed agent replays the motion it executed in the first run. The
two runs differ only in the instructions; what the instructed agents
visibly do is held fixed.

\begin{proposition}[Soundness]
\label{prop:sound}
If a generator is causally faithful, every uninstructed agent executes
identical motion in the two runs.
\end{proposition}

\noindent\emph{Proof sketch.}
Both runs start from the same $c^0$. Suppose they agree on $c^t$. An
uninstructed agent has $C_i=\emptyset$ in both runs, so by
Eq.~\eqref{eq:faithful} its next motion is determined by $c^t$ and
$\xi$ and is the same in both. Since the instructed agents replay the
same motion, the runs agree on $c^{t+\Delta}$, and by induction at
every step.
\qed

A faithful generator therefore scores exactly zero, and any difference
is intent leakage: \emph{an effect of the instructions on other agents
that did not pass through executed motion}. For an uninstructed agent $i$
with positions $x_i^t$ and $\tilde x_i^t$ in the two runs, we measure
the displacement
\begin{equation}
d_i^t = \big\|x_i^t-\tilde x_i^t\big\|
\label{eq:displacement}
\end{equation}
and report the share of agents it affects and its magnitude over the
rollout (Sec.~\ref{sec:experiments}). The test isolates leakage through any generator supporting fixed randomness and trajectory replay, instruction privacy is addressed separately in Sec.~\ref{sec:data}.

\begin{figure*}[t]
\centering
\includegraphics[width=\textwidth]{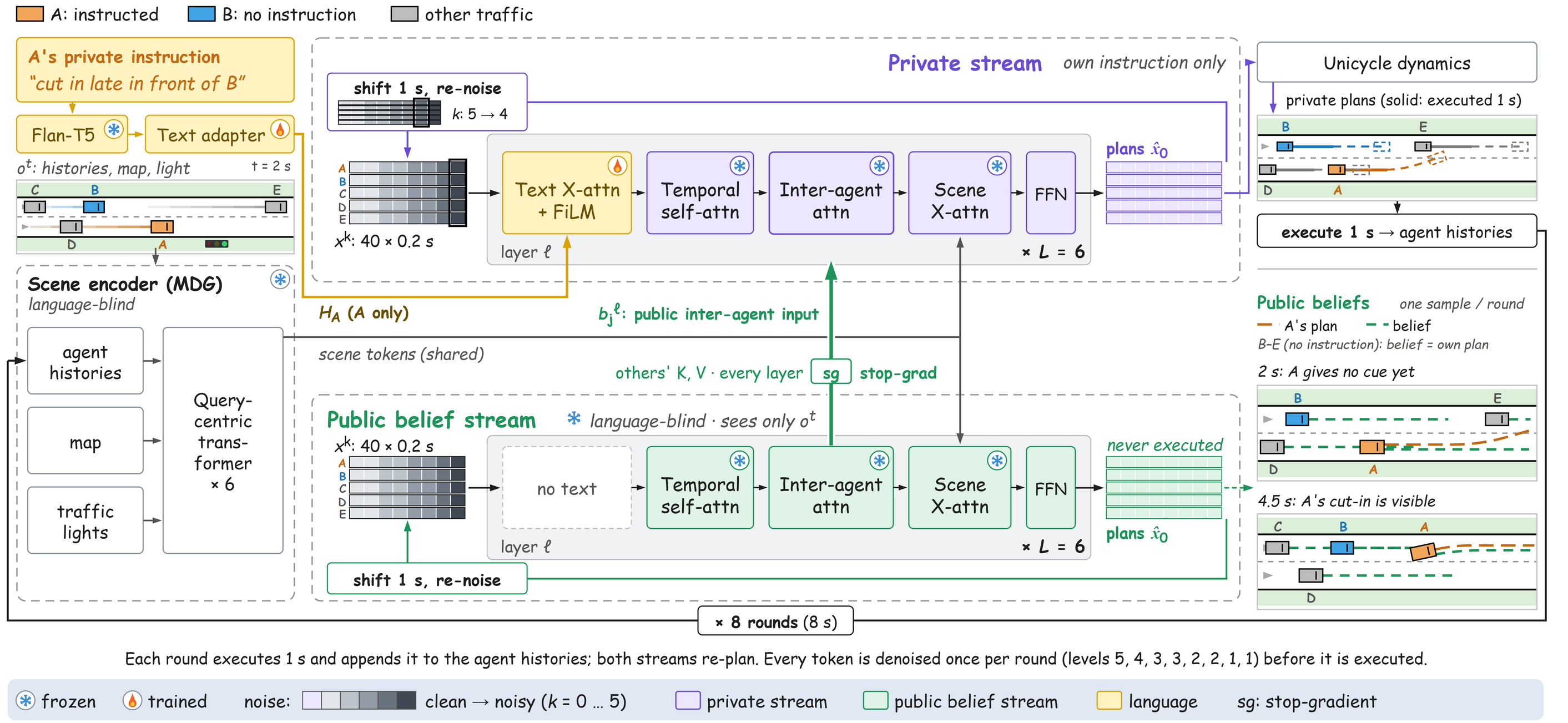}
\caption{\textbf{CaFe.}
The private stream conditions each agent on its own instruction,
while a language-blind pass of the frozen backbone predicts public
beliefs from committed motion. In inter-agent attention, an agent
reads its own private features and the public beliefs of others,
never their private futures or instructions. After each one-second
commit, both streams update from the new observable history. The
example on the right shows how public beliefs change as an agent's
motion reveals its intention.}
\label{fig:arch}
\end{figure*}

\section{Method: CaFe}
\label{sec:method}

CaFe separates what an agent intends from what others can observe
(\cref{fig:arch}). A \emph{private stream} plans each agent's motion
from the context and its own instruction. A \emph{public stream} plans
the same scene without any instruction. Its predictions are the
\emph{beliefs} that any observer could form from executed motion. The
two streams meet in one place: in inter-agent attention, an agent
reads its own private features and the public beliefs about everyone
else. Agents still anticipate each other, but only through what has
been executed, never through another agent's instruction or plan.


\subsection{Diffusion-forcing Backbone in Closed-Loop}
\label{sec:backbone}
We build on MDG~\cite{huang2025mdg}, a multi-agent simulator based on
Diffusion Forcing~\cite{chen2024diffusionforcing}. A scene encoder
embeds the agent histories, map elements and traffic signals. The
denoiser stacks temporal self-attention, inter-agent attention and
scene cross-attention over the agents' future action tokens
(acceleration and yaw rate), which differentiable unicycle dynamics
integrate into trajectories. Every token carries its own noise level,
and masked denoising trains the model to recover clean actions from
any mixture of levels. We keep this architecture and objective and
refer to~\cite{huang2025mdg} for details.

Per-token noise levels make closed-loop generation coherent. Every
$\Delta=1$\,s the model executes the first second of its 8\,s plan,
$c^t\rightarrow c^{t+\Delta}$, then shifts the remainder forward and
re-noises it, lightly for the near future and more strongly further
ahead. A single denoising pass gives the next plan, so near-term
motion stays consistent while distant motion can still respond to
what other agents execute.



\subsection{Dual-stream denoising}
\label{sec:dual}

At every re-plan, CaFe runs the frozen backbone twice over the same
context $c^t$ and the same sampling noise: a public pass without language and a private pass with it. Each stream reuses its own previous plan, and only the private plan is executed.

\paragraph{Private stream.}
A frozen text encoder followed by a small trainable adapter maps the
instruction $C_i$ to features $H_i$. Before every denoiser layer, each token $h_{i,k}$ of agent $i$ cross-attends to $H_i$ and is modulated by FiLM~\cite{perez2018film}, as in per-agent text adapters on frozen
simulators~\cite{chen2026ecosim}:
\begin{equation}
\begin{aligned}
(\Delta\gamma_{i,k},\beta_{i,k}) &=
g\big(\mathrm{CrossAttn}(h_{i,k},H_i)\big),\\
h_{i,k} &\leftarrow
(1+\Delta\gamma_{i,k})\odot h_{i,k}+\beta_{i,k}.
\end{aligned}
\label{eq:film}
\end{equation}
The output of $g$ is initialised to zero, so training starts from the
backbone. Agents without an instruction skip the module, and the
scene encoder never sees text, so no instruction enters the shared
scene features.


\paragraph{Public stream.}
The public stream reuses the language-blind backbone to predict
other agents from observed motion. This retains learned anticipation
without introducing a separate belief model. Its tokens $b_{j,k}$
therefore depend only on the context and instruction-independent
randomness,
\begin{equation}
b_{j,k} = G_{j,k}\big(c^t,\xi\big),
\label{eq:belief}
\end{equation}
a function of the context and of randomness that carries no
instruction. Since the context updates with every executed second, the
beliefs follow what agents reveal by moving (\cref{fig:arch}, right),
although the public plan itself is never executed. For an agent
without an instruction, the private pass has the same inputs as the
public one, noise included, so its tokens coincide in the two streams
even when other agents are instructed. Without any instruction, CaFe reduces exactly to the backbone.


\noindent\textbf{Belief attention.} In the backbone, inter-agent attention lets every agent read the
future tokens of all others. Once those tokens are shaped by language,
they become a path for intent leakage. CaFe keeps the query private
and takes the keys and values of the other agents from their beliefs.
For token $k$ of agent $i$, the attention inputs are
\begin{equation}
\begin{aligned}
Q &= h_{i,k},\\
K = V &= \{h_{i,k}\}\cup
\{\operatorname{stop\_grad}(b_{j,k})\}_{j\ne i},
\end{aligned}
\label{eq:belief_attn}
\end{equation}
where $\operatorname{stop\_grad}$ blocks gradients into the public
pass. The layer's projections and relation encodings are unchanged.
The streams thus meet only here, and only in one direction: private
tokens read beliefs, and nothing reads private tokens but their own
agent.



\subsection{Causal faithfulness}
\label{sec:cafe_faithful}

\begin{proposition}[Leak-freedom by construction]
\label{prop:leakfree}
CaFe is causally faithful.
\end{proposition}

\noindent\emph{Proof sketch.}
Fix the context $c^t$ and the randomness $\xi$. The scene features
and the beliefs are computed without text (Eq.~\eqref{eq:belief}), so
an instruction enters only the private tokens of its own agent. No
other agent reads these tokens (Eq.~\eqref{eq:belief_attn}), and the
plan they produce is reused by that agent alone. By induction over
re-plans, $x_i^{t+\Delta}$ is therefore a function of $c^t$, $\xi$
and $C_i$, which is Eq.~\eqref{eq:faithful}.
\qed

The paired test of Sec.~\ref{sec:metric} must thus measure exactly
zero for CaFe (Proposition~\ref{prop:sound}), which
Sec.~\ref{sec:experiments} confirms. This holds under test-time
guidance~\cite{huang2025mdg} as well, provided the guidance respects
the same interface. Guidance adjusts plans that leave the road or
overlap another agent's plan, and taking the others' private plans as
obstacles would reopen the path that belief attention closes. CaFe
instead checks each private plan against the others' public plans,
which carry no instruction.



\subsection{Closed-loop training}
\label{sec:training}

CaFe is trained the way it is run: in closed loop. Masked denoising
learns from logged histories, whereas at test time the model plans
from motion it executed itself and from plans it has re-noised. We
close this gap with \emph{closed-loop fine-tuning with the energy
score} (CLFT-ES). Each scene is rolled out $M$ times over eight
one-second re-plans with the test-time sampler, and gradients flow
back through the executed states and the reused plans, so an early
re-plan is held responsible for what it causes later. Since the log
shows only one of many plausible futures, the rollouts are scored
against it as a set:
\begingroup
\setlength{\abovedisplayskip}{4pt}
\setlength{\belowdisplayskip}{4pt}
\begin{equation}
\begin{aligned}
\mathrm{ES}_i
={}& \frac{1}{M}\sum_m
\lVert \tau_i^{(m)}-y_i\rVert \\
&-\frac{1}{2M(M-1)}\sum_{m\neq m'}
\lVert \tau_i^{(m)}-\tau_i^{(m')}\rVert,
\end{aligned}
\label{eq:clft_es}
\end{equation}
\endgroup
where $\tau_i^{(m)}$ is agent $i$'s motion executed so far in rollout 
$m$ followed by its current plan. $y_i$ the logged trajectory and 
$\lVert\cdot\rVert$ an RMS distance over position, heading and speed. 
The first term pulls the rollouts towards the log and the second keeps 
them apart to avoid mode collapse. The score is averaged over all 
re-plans, so every plan is supervised, not only the executed outcome.

CLFT-ES trains both the backbone and CaFe. The backbone is first fine-tuned
without language and then frozen. The language modules of
Eq.~\eqref{eq:film} are trained on \dataname{} with the
dual-stream model in the loop. Instructions are thus learned in the
setting they are tested in: over full rollouts, with every other
agent reading beliefs. Throughout, half of the scenes have most
agents execute their logged motion while their own plans are still
reused and scored. Their history then contradicts what they planned,
which is the situation of the public stream at every re-plan: it
reads executed private motion but reuses its own earlier belief.
Whole instructions and single clauses are also dropped at random, so
the model handles partial and missing instructions
(Sec.~\ref{sec:language}).

\begin{algorithm}[t]
\caption{Closed-loop fine-tuning with energy score (CLFT-ES).}
\label{alg:clft_es}
\footnotesize
\begin{algorithmic}[1]
\Statex \textbf{Input} Model $\theta$, logged trajectories $\{y_i\}$,
instructions $C$, $M\geq 2$
\State $\mathcal R \gets \operatorname{ReplayMask}()$
\Comment{most agents in half the scenes}
\For{$m=1,\ldots,M$}
    \State $c^{0}\gets$ logged context, $r^{0}\gets\varnothing$
    \For{$t=0,\Delta,\ldots,7\Delta$}
        \State $\hat x^{t}\gets
        \operatorname{Plan}_{\theta}(c^{t},\mathcal{C},r^{t},\xi^{t})$
        \State $\tau_i^{t,(m)}\gets x_i^{(0,t]}\oplus\hat x_i^{t}$
        for all $i$
        \Comment{executed, then planned}
        \State $x^{t+\Delta}\gets
        \operatorname{Execute}_{\Delta}(\hat x^{t})$
        \State $x_j^{t+\Delta}\gets y_j^{t+\Delta}$
        for $j\in\mathcal R$
        \State $c^{t+\Delta}\gets
        \operatorname{Update}(c^{t},x^{t+\Delta})$
        \State $r^{t+\Delta}\gets
        \operatorname{ReNoise}(\operatorname{Shift}(\hat x^{t}))$
    \EndFor
\EndFor
\State $\mathcal L_{\mathrm{ES}}\gets
\operatorname{mean}_{i,t}\,
\mathrm{ES}_i\big(\{\tau_i^{t,(m)}\}_{m=1}^{M},y_i\big)$
\Comment{Eq.~\eqref{eq:clft_es}}
\State $\mathcal L\gets
\mathcal L_{\mathrm{ES}}+\mathcal L_{\mathrm{aux}}$
\Comment{masked denoising and non-collision}
\State $\theta\gets
\operatorname{OptStep}(\theta,\nabla_\theta\mathcal L)$
\end{algorithmic}
\end{algorithm}

\section{\dataname{} for Private, Verifiable Instructions}
\label{sec:data}

Training and evaluation require four data properties:
\textbf{verifiable facts}, \textbf{private instructions},
\textbf{symmetric coverage} across vehicles, and
\textbf{reaction annotations} for selective prompting.
While the replay test reports leakage through models, instruction
privacy must be ensured during data construction.
ProSim-Instruct-520k and InterDrive include
scene-level or cross-agent descriptions in single-sentence instructions~\cite{tan2024prosim,chang2025langtraj},
which can expose information beyond a recipient's observed history.
WOMD-PACT instead derives private instructions from measured behaviour
on all 531,099 WOMD training and validation
scenes~\cite{ettinger2021womd}, each spanning 9.1\,s at 10\,Hz.
It comprises timed behaviour tags, an interaction graph and private
language instructions.

\subsection{Stage 1. Action Tag}
\label{sec:tags}

The first stage turns each track into atomic events with a start and
an end time. For stops, turns and lane changes, a strict threshold
detects the action, and hysteresis determines its start and end.

Speed tags include \texttt{Accelerate}, \texttt{Decelerate},
\texttt{MaintainSpeed} and \texttt{Stationary}, as well as
\texttt{Reverse} and \texttt{Parked} for vehicles. Vehicles and
cyclists receive \texttt{TurnLeft} and \texttt{TurnRight} tags.
For vehicles, \texttt{LaneShiftLeft} and \texttt{LaneShiftRight}
cover lane changes, splits and merges. We measure lateral motion
relative to the connected lane centrelines along the vehicle's
route to distinguish lane shifts from bends in the road.
Pedestrians receive \texttt{CrossesVehiclePath} where applicable.
We omit pedestrian turn tags because their bounding-box heading
can differ from their direction of travel. Tracks that are too
short or fragmented to label receive an explicit status, distinct 
from the absence of tags.

A scene contains on average 68 agents, of which about 55 receive
tags. Of those, 19 do more than remain parked or maintain speed,
contributing around 30 tags per scene. The next stage identifies
which actions occur in response to another agent.





\subsection{Stage 2. Interaction graph}
\label{sec:graph}

The second stage determines what each reaction responds to. A
reaction is a vehicle's \texttt{Decelerate}, \texttt{Stationary} or
lane-change event. For each reaction we launch \emph{ghosts} of the
vehicle at its start and every second until it ends. Each ghost
leaves from the vehicle's actual state, accelerates to its speed
before the reaction, or to its later cruising speed if it was
waiting, and follows its path as if it had not reacted. The first
agent a ghost meets is a parent, provided it reaches the conflict
point first and in its recorded track. If the ghost reaches a red
stop line first, the reaction is attributed to the signal.

Each attributed reaction forms an \emph{episode}, and episodes are
linked when one agent's wait depends on another's reaction, forming
chains such as a queue behind a vehicle yielding to a pedestrian.
Unlike proximity or arrival-order
labelling~\cite{sun2022m2i,kumar2021trafficgraph}, the ghost test
rejects agents that merely pass nearby. A scene contains on average
10.2 episodes, 1.6 of them held by a red light, and 5.6 links.
Details and validation are in the supplementary material.




\begin{table*}[!t]
\centering

\begingroup
\small
\setlength{\tabcolsep}{3.4pt}
\renewcommand{\arraystretch}{1.0}

\begin{tabular}{@{}l ccc ccc cccc ccc@{}}
\toprule
& \multicolumn{6}{c}{Leakage}
& \multicolumn{4}{c}{Realism$\uparrow$}
& \multicolumn{3}{c}{Accuracy and safety$\downarrow$} \\
\cmidrule(lr){2-7}
\cmidrule(lr){8-11}
\cmidrule(l){12-14}
\multirow{2}{*}{Method}
& \multicolumn{3}{c}{Affected (\%)$\downarrow$}
& \multicolumn{3}{c}{Displacement (m)$\downarrow$}
& \multirow{2}{*}{Meta} & \multirow{2}{*}{Kin.} & \multirow{2}{*}{Inter.} & \multirow{2}{*}{Map}
& \multirow{2}{*}{minADE} & \multirow{2}{*}{Coll. (\%)} & \multirow{2}{*}{Off-road (\%)} \\
\cmidrule(lr){2-4}
\cmidrule(lr){5-7}
& 1\,s & 4\,s & 8\,s
& 1\,s & 4\,s & 8\,s
& & & & & & & \\
\midrule
ECoSim~\cite{chen2026ecosim}
& 0.1 & 7.1 & 22.6
& 1.54 & 1.46 & 3.08
& 0.7685 & 0.4642 & 0.8032 & 0.8977
& 1.507 & 3.49 & 11.54 \\
\midrule
Backbone (no text)
& n/a & n/a & n/a
& n/a & n/a & n/a
& 0.7599 & 0.4493 & \textbf{0.7935} & 0.8940
& 1.417 & \textbf{4.34} & \textbf{11.68} \\
\midrule
Inter-agent attention~\cite{huang2025mdg,chang2025langtraj, chen2026ecosim}
& 0.1 & 11.2 & 26.7
& 0.78 & 1.40 & 2.77
& 0.7628 & 0.4624 & 0.7903 & 0.8990
& 1.362 & 5.82 & 12.22 \\
Broadcast instructions~\cite{tan2024prosim,chang2025langtraj}
& 22.2 & 69.2 & 82.7
& 0.87 & 2.32 & 4.14
& \textbf{0.7651} & 0.4694 & 0.7917 & \textbf{0.9000}
& 1.435 & 6.06 & 12.18 \\
History only
& \textbf{0.0} & \textbf{0.0} & \textbf{0.0}
& n/a & n/a & n/a
& 0.7475 & \textbf{0.4744} & 0.7608 & 0.8863
& 1.654 & 10.42 & 14.28 \\
\rowcolor{cafeblue}
\textbf{\method{} (ours)}
& \textbf{0.0} & \textbf{0.0} & \textbf{0.0}
& n/a & n/a & n/a
& 0.7631 & 0.4602 & \textbf{0.7935} & 0.8972
& \textbf{1.276} & 4.81 & 12.10 \\
\bottomrule
\end{tabular}
\endgroup

\caption{\textbf{Intent leakage and realism} on 1,000 \dataname{}
validation scenes with 32 rollouts each.
Variants reproduce the information routes of the cited works on
a shared backbone with identical training data, sampler and
leak-free guidance, isolating what each agent can observe.
ECoSim~\cite{chen2026ecosim} is evaluated as published as an
external autoregressive baseline.
\textit{Affected} is the percentage of nearby moving uninstructed
agents displaced by more than 0.5\,m when instructed agents replay
identical motion with their instructions removed, evaluated on
the 746 eligible scenes (750 for ECoSim).
\textit{Displacement} is the mean displacement among affected agents.
Realism and its components follow WOSAC 2025. 
\textit{n/a} means that the leakage test is inapplicable
(no-text backbone) or that mean displacement is undefined
because no agents were affected.
Bold indicates the best result among variants sharing our backbone.}
\label{tab:main}

\end{table*}


\subsection{Stage 3. Language}
\label{sec:language}

The events of every vehicle present at $t=0$ that does more than stay
parked or hold its speed are grouped into clauses. A clause joins a
speed event with the course events it overlaps, whereas a reaction
always forms its own clause. Reactions are events the graph attributes
to another agent or a red light. All others remain unclassified
because no test establishes a driver's own choice. Each clause states its
manoeuvre and direction, while timing, intensity, amount and speed or
angle are each stated in half of the clauses. \dataname{} comprises
1.49 million clauses for 0.81 million vehicles in 70,471 scenes,
30.4\% of them reactions.

\noindent\textbf{Private.}
A clause describes only its own vehicle, never other road users or
traffic signals, and a reaction clause never states its timing. Up to
two map facts, such as the lane position or a nearby junction,
crosswalk or stop sign, are added at random rather than by behaviour,
so that a mention does not predict it.

\noindent\textbf{Verifiable.}
Gemma-4-26B-A4B-it, served with vLLM~\cite{kwon2023vllm}, writes each
clause as a sentence. Mechanical checks and a separate call that lists
the behaviours in the sentence must agree with the clause.
Within three attempts, 95.1\% of clauses pass, and the rest are dropped.

\noindent\textbf{Dropout.}
Training drops whole instructions and single clauses with probability
0.3 each. At test time, every instructed vehicle receives all its
clauses.





\section{Experiments}
\label{sec:experiments}





\subsection{Setup}
\label{sec:experiment_setup}



We evaluate on the \dataname{} validation split. The scene-level comparison uses 1,000 scenes with 32 rollouts each. For targeted evaluation, we consider 500 interactions between instructed agents $A$ and uninstructed responders $B$, using eight matched rollouts across methods. Instruction satisfaction and responder behaviour are evaluated on the corresponding eligible interactions. Realism and guidance are run with 32 rollouts each. We report paired 95\% confidence intervals for selected comparisons.

\par\noindent\textbf{Metrics.}
The paired test of \cref{sec:metric} replays executed motion
without instructions under matched noise and counts nearby moving
agents (within 30\,m) displaced by more than 0.5\,m. All instructed
agents are replayed in \cref{tab:main}, and $A$ alone, with its
linked responders, in the targeted interactions. With $A$ replayed,
$B$ reacts if it slows by at least 0.5\,m/s within 3\,s of the
logged reaction, and we count collisions between $A$ and $B$. In
free rollouts, a clause is satisfied when its tagged behaviour
begins within 1\,s of its annotated start time $s_k$. Realism,
collision and off-road rates follow WOSAC
2025~\cite{montali2023wosac}, and minADE keeps the best of 32
rollouts.

\par\noindent\textbf{Variants.}
To isolate information access from differences in data and
architecture, we reproduce published information routes within
a shared backbone. \emph{Broadcast instructions} shares text
across agents~\cite{tan2024prosim,chang2025langtraj}.
\emph{Single stream} exposes other agents' private future
tokens~\cite{huang2026vbd,huang2025mdg}, labelled
\emph{inter-agent attention} in \cref{tab:main}.
\emph{History only} removes inter-agent attention from the
denoiser~\cite{wu2024smart}, whereas \method{} uses public
beliefs from the frozen backbone (\cref{sec:backbone}).
All variants share training data and sampler. The no-language
backbone serves as a reference, and ECoSim~\cite{chen2026ecosim}
as an external baseline.

\subsection{Leakage through text and future plans}

In the scene-level paired test (\cref{tab:main}), broadcast
instructions affect 82.7\% of nearby uninstructed agents at
8\,s. Access to others' private future tokens affects 26.7\%
despite per-agent instructions, while ECoSim affects 22.6\%.
Per-agent prompting therefore does not prevent leakage.
\method{} records 0.0\% affected agents while maintaining
WOSAC realism comparable to single stream (0.7631 versus
0.7628). Among variants sharing our backbone, it achieves
the lowest minADE (1.276\,m versus 1.362\,m for single stream)
and reduces the overall collision rate from 5.82\% to 4.81\%.
These results show that enforcing the information boundary
need not compromise trajectory fidelity or simulation quality.

In targeted replays, single stream still displaces nearby uninstructed agents despite identical executed motion from A.
\method{} records 0.0\% in both scene-level and targeted tests,
consistent with Proposition~\ref{prop:leakfree}.

\subsection{Effect of public belief modelling}

Both history only and \method{} prevent leakage
(\cref{tab:main}), allowing us to examine the value of learned
public beliefs within the same information boundary.
Both use executed motion, but \method{} additionally predicts
other agents' behaviour through the text-free frozen backbone.
In fixed-motion replay, $B$ reacts in 81.7\% of \method{}
rollouts versus 79.5\% for history only, while collisions with
$A$ fall from 9.8\% to 5.8\% ($[-6.8,-1.5]$\,pp).
WOSAC realism also improves from 0.7238 to 0.7481
(\cref{tab:response_quality}). These results indicate that
learned public beliefs support anticipation and reactive
behaviour without exposing private plans.



\begin{table}[!t]
\centering

\caption{\textbf{Instruction success and replay response quality.}
Clause satisfaction and responder metrics use eight matched
rollouts per item. Responder metrics use fixed-motion replay
with unprompted $B$.}
\label{tab:response_quality}
\label{tab:attention_ablation}

\begingroup
\small
\setlength{\tabcolsep}{3pt}
\renewcommand{\arraystretch}{1.08}

\begin{tabular*}{\columnwidth}{
@{\extracolsep{\fill}}lcccc@{}}
\toprule
Method
& \shortstack{Clause\\(\%)$\uparrow$}
& \shortstack{Reacts\\(\%)$\uparrow$}
& \shortstack{$B$ hits $A$\\(\%)$\downarrow$}
& \shortstack{Meta$\uparrow$\\(guidance-free)} \\
\midrule

No language
& 38.0 & 81.7 & 5.8 & \textbf{0.7494} \\

Single stream
& \textbf{40.9} & \textbf{82.4}
& \textbf{5.3} & 0.7466 \\

History only
& 35.8 & 79.5 & 9.8 & 0.7238 \\

\midrule
\rowcolor{cafeblue}
\textbf{\method{} (ours)}
& 40.8 & 81.7 & 5.8 & 0.7481 \\

\bottomrule
\end{tabular*}
\endgroup
\end{table}

\begin{table}[!t]
\centering
\caption{\textbf{Effect of guidance.}
WOSAC 2025 realism, collision and off-road rates
without and with guidance.
Guided \method{} has 0.0\% affected agents at 8\,s.}
\label{tab:guidance_ablation}

\begingroup
\small
\setlength{\tabcolsep}{3pt}
\renewcommand{\arraystretch}{1.05}

\begin{tabular*}{\columnwidth}
{@{\extracolsep{\fill}}llccc@{}}
\toprule
Method & Guide
& Meta$\uparrow$
& Coll.$\downarrow$
& Off-road$\downarrow$ \\
\midrule

Backbone
& No
& 0.7494
&  5.6
& 13.7 \\
& Yes
& 0.7599
& \textbf{4.34}
& \textbf{11.68} \\

\addlinespace[2pt]
Single stream
& No
& 0.7466
& 8.8
& 14.8 \\
& Yes
& 0.7628
& 5.82
& 12.22 \\

\addlinespace[2pt]
\rowcolor{cafeblue}
\textbf{\method (Ours)}
& No
& 0.7481
& 7.1
& 14.5 \\
\rowcolor{cafeblue}
& Yes
& \textbf{0.7631}
& 4.81
& 12.10 \\

\bottomrule
\end{tabular*}
\endgroup
\end{table}

Guidance~\cite{huang2025mdg} improves all models (\cref{tab:guidance_ablation}).
For \method{}, it raises the WOSAC meta score from 0.7481 to
0.7631 and lowers collisions from 7.1\% to 4.81\%. Guidance
checks each private plan against public information rather than private
plans of others, retaining 0.0\% measured leakage.

\subsection{Instruction following and interaction quality}

We next assess whether enforcing this information boundary
preserves instruction following and reactive behaviour.
Instruction following is evaluated in free generation,
and responses under fixed-motion replay.
\method{} satisfies 40.8\% of timed clauses, compared with
40.9\% for single stream ($[-2.1,+1.7]$\,pp) and 38.0\%
without language. With $A$ replayed, $B$ reacts in 81.7\%
of \method{} rollouts versus 82.4\% for single stream,
and collides with $A$ in 5.8\% versus 5.3\%
($[-0.2,+1.3]$\,pp). WOSAC realism remains close to the
no-language backbone (0.7481 versus 0.7494,
\cref{tab:response_quality}). Thus, \method{} removes measured
leakage while retaining comparable performance on these measures.

\noindent\textbf{End-to-end interaction.}
We then evaluate instruction following and response jointly,
with all agents simulated and only $A$ instructed.
Success requires $A$ to perform its behaviour at the annotated
time, $B$ to react within 3\,s of the logged human reaction,
and no collision between them. \method{} achieves 32.4\%
joint success versus 33.9\% for single stream. Among rollouts
in which $A$ performs its manoeuvre, $B$ responds without
collision in 83.5\%.

\noindent\textbf{Causal order.}
Joint success does not establish whether responses follow
observable changes in $A$'s motion. In a separate paired
free-rollout test, we vary $A$'s private instruction under
matched noise and record when each agent's executed motion
first differs from its uninstructed rollout.
With \method{}, $B$ never departs before $A$ does,
whereas with single stream, $B$ departs first in 2.6\%
of rollouts. This complements fixed-motion replay by
examining the temporal order of instruction-induced changes
when all agents evolve freely.

\section{Conclusion}
\label{sec:conclusion}

We formalize intent leakage and introduce a paired replay test
that separates it from legitimate reactions to observed motion.
\dataname{} addresses privacy in instruction construction, while
\method{} enforces it in inter-agent information flow.
Our results show that eliminating measured leakage is compatible
with comparable realism, instruction following and response
quality. This establishes the feasibility of causally faithful
conditional simulation and motivates stronger leak-free
simulators. Extending this principle across conditioning modalities
would broaden its use in controllable scenario generation.

\par\noindent\textbf{Limitations.} 
Our evaluation focuses on language conditioning. The same
information boundary can be formulated for goal points, route
sketches and other private conditions, but these extensions
remain untested. \method{} demonstrates one practical implementation
using a frozen backbone to form public beliefs. More elaborate
belief representations and conditioning mechanisms remain to be
explored, offering scope to improve generation quality while
preserving causal faithfulness.

{
    \small
    \bibliographystyle{ieeenat_fullname}
    \bibliography{refs}
}

\end{document}